# Reducing Quantum Cost in Reversible Toffoli Circuits

Marek Szyprowski[1]
[1]Institute of Computer Science
Warsaw University of Technology
Nowowiejska 15/19, 00-665 Warsaw, Poland
m.szyprowski@ii.pw.edu.pl

Pawel Kerntopf[1,2]
[2]Department of Theoretical Physics and Informatics
University of Łódź
Pomorska 149/153, 90-236 Łódź, Poland
p.kerntopf@ii.pw.edu.pl

*Abstract*— **Recently, reversible circuit synthesis has been intensively studied. One of the problems that has not been solved for a long time was exact minimization of gate count (GC) in 4-bit circuits. Finally, last year a tool of practical usage for finding optimal gate count Toffoli networks for any 4-variable function was developed. However, not much work has been done yet on exact minimization of quantum cost (QC) in 4-bit circuits. This paper presents an application of the above mentioned tool to reducing QC of 4-bit reversible circuits. It is shown that for benchmarks and for designs taken from recent publications it is possible to obtain savings in QC of up to 74% comparing with previously known circuits.**

## I. Introduction

A gate (circuit) is called reversible if there is a one-to-one correspondence between its inputs and outputs. Research on reversible logic circuits is motivated by advances in quantum computing, nanotechnology and low-power design. Therefore, reversible logic synthesis has been intensively studied. The attention has been focused on the synthesis of circuits built from the NCT library of gates.

Satisfactory reversible logic synthesis algorithms for arbitrary libraries of gates and cost functions have not been found yet. In addition, NCT library synthesis techniques developed for such circuits scale not well and optimal circuits not always can be found even for relatively small numbers of inputs and outputs. Many reversible circuit synthesis algorithms have been proposed. However, they sometimes generate very suboptimal circuits. Therefore, to improve circuit quality, different techniques, consisting of a series of local optimizations have been developed. However, the problem is that some circuits cannot be reduced only by local optimization. Also, for larger functions these algorithms require many iterations of local optimization that leads to a high runtime.

For many years only few exact optimal circuits have been found for *n*-variable functions with $n > 3$. For example, it was even not known what is the maximal gate count in optimal circuits implementing 4-bit reversible functions. Last year a very fast tool capable of synthesizing optimal circuits for any 4-bit reversible specification was finally developed [2] what has become a real breakthrough. With this tool it was possible to establish that there are 144 4-bit functions requiring 15 gates in their optimal circuits and that there exists none requiring 16 gates [3, 12].

Quality of a reversible circuit is usually estimated by gate count (GC) or by a metric called quantum cost (QC). Much less effort has been devoted to minimization of QC in reversible circuits. Maslov et al. [14] used a mixture of different techniques (including MMD algorithm, Reed-Muller spectra based algorithm, template application and resynthesis) to improve either gate count or quantum cost what led to improving results for some benchmarks from [12]. However, exact minimization of QC was not the aim of this approach. Donald and Jha [1] added a new option for optimizing QC to their earlier algorithm. Due to this they were able to improve results for some benchmarks but have attempted to find exact minimal values for small circuits only. They performed similar experiments for an extended library of gates including also SWAP, Fredkin and Peres gates. Wille et al. [27] formulated a synthesis problem as a quantified Boolean formula and then solved it by applying Binary Decision Diagrams. This enabled to find the minimal as well as the maximal QCs for the specified number of gates up to seven gates. However, these calculations were performed only for minimal gate count circuits. Grosse et al. [4] considered synthesis for networks made of multiple-control Toffoli gates using SAT-like engines. Their approach to reducing quantum cost was the same, i.e. minimization of the number of gates as the first step and only then trying to reduce quantum cost with the fixed gate count. However, as we will show in the paper, finding exact minimal quantum cost circuits for larger functions requires considering circuits having greater number of gates than the minimal size ones.

Some efforts have been recently made to reduce QCs of designs. One of the recent approaches consists in looking for circuit realizations using quantum elementary gates like NOT, controlled NOT, two square roots of NOT [16, 17, 20, 21] or Hadamard gates [15]. However, we show that significant reduction of QC can also be obtained without considering elementary quantum gate library.

First, we have developed a tool similar to the one reported earlier in [2] capable of finding a circuit having a minimal number of gates for any 4-bit reversible function. Then we have programmed and run an approach capable of finding *all* circuits implementing a given reversible function *with a specified value of GC.* Our calculations have led to obtaining savings in QC over 50% for some known benchmark functions and for designs reported in recent publications on synthesis of reversible circuits. At the same time we have shown that the numbers of *all* quantum cost minimal circuits for larger designs achieve the order of $10^7$.

The paper is organized as follows. Section II recalls basic concepts of reversible logic. In Section III notions of cost functions and optimal reversible circuits are introduced. Sections IV presents a survey of recent results on synthesis of 3-bit and 4-bit reversible Boolean circuits. In Section V details of the fast algorithm for optimal gate count synthesis of 4-bit circuits are overviewed. Section VI contains description of our approach to searching for minimal quantum cost 4-bit reversible circuits. In Section VII our experimental results are collected and compared to known circuits from benchmark pages and from the literature. Section VIII summarizes the paper with conclusions and suggestion for further research.

## II. PRELIMINARIES

**Definition 1.** A completely specified *n*-input *n*-output Boolean function (referred to as *n**n function) is called *reversible* if it maps each input assignment into a unique output assignment.

There are $2^n!$ reversible *n**n Boolean functions. For $n = 3$ this number is equal to 40,320 and for $n = 4$ is greater than $2{\cdot}10^{13}$.

**Definition 2.** An *n*-input *n*-output (*n**n) gate (or circuit) is *reversible* if it realizes an *n**n reversible function.

In a reversible circuit fanout of each gate output is always equal to 1. As a consequence *n**n reversible circuits can be only build as a cascade of *k**k reversible gates ($k \le n$).

**Definition 3.** A set of reversible gates that can be used to build reversible circuits is called a *gate library*.

Many gate libraries have been examined in the literature. The so called *NCT library* consists of 1*1 NOT, 2*2 CNOT and 3*3 TOFFOLI gates. In this paper we will also discuss the usage of 4*4 TOFFOLI gates and 3*3 PERES gates. Below we define all these gates.

**Definition 4.**

1*1 ***NOT***(*x*) *gate* performs the operation

$$(x) \to (x \oplus 1),$$

2*2 ***CNOT***(*x,y*) *gate* performs the operation

$$(x, y) \to (x, x \oplus y),$$

3*3 ***TOFFOLI***(*x,y,z*) *gate* performs the operation

$$(x, y, z) \to (x, y, z \oplus xy),$$

4*4 ***TOFFOLI***(*x,y,z,u*) *gate* performs the operation

$$(x, y, z, u) \to (x, y, z, u \oplus xyz),$$

3*3 ***PERES***(*x,y,z*) *gate* performs the operation

$$(x, y, z) \to (x, x \oplus y, z \oplus xy),$$

where ⊕ denotes XOR operation.

The first four of the above defined gates (denoted N, C, T, T4, respectively) invert one input if and only if all others are 1, passing the other inputs unchanged to corresponding outputs. Each of the N, C, T, T4 gates is invertible, i.e. equal to its own inverse. Let us note that a PERES(*x,y,z*) gate is equivalent to a *TOFFOLI*(*x,y,z*) gate followed by a *CNOT*(*x,y*) gate. *Inverted PERES* gate is not equal to PERES gate. We will denote them by $P^{-1}$ and P, respectively [13].

## III. COST FUNCTIONS AND OPTIMAL REVERSIBLE CIRCUITS

For any reversible function there exist many reversible circuits implementing it. Thus a cost function has to be defined to evaluate the quality of a circuit. Usually, additive cost functions are applied. Therefore, adding a gate to a circuit leads to increasing its cost. A cost of each gate type is often expressed as a non-zero value. The cost of the (hypothetically) simplest gate is assumed to be 1.

A number of gate cost functions have been proposed (see, e.g. [28]). The simplest cost function of a reversible circuit is equal to total number of gates. It is called *gate count* (GC in short). Although this cost function probably would not be applicable to future technologies, it is used in the literature for comparing the quality of synthesis algorithms. Nevertheless, it could be treated as an approximation of practical cost functions for small reversible circuits.

Other cost functions are also considered. The most widely used, called *quantum cost* (QC), is based on cost of elementary quantum gates as each reversible gate can be built from several elementary quantum gates. It is assumed that the cost of each elementary quantum gate equals 1, so the cost of a reversible gate equals to the total number of elementary quantum gates used. The quantum cost of N, C, P, T and T4 gates is assumed to be 1, 1, 4, 5 and 13, respectively (see for example [5, 12]).

By an *optimal circuit* we mean the best possible implementation, i.e. the one having the minimal cost. The set of optimal circuits implementing a reversible function depends on a gate library and a cost function. It should be noted that a reversible function may have many equivalent optimal circuits. A database of all optimal circuits can be generated by building recursively new circuits from all gates that belong to the gate library and selecting only the circuits with the minimal cost after incrementing the number of gates in the circuits by 1. Such databases have been described in the literature, but usually only one optimal circuit is stored for each function (see, e.g. [14, 18].

Let us define an equivalence class in the set of reversible functions with respect to the cost of optimal circuits realizing these functions.

**Definition 5.** Two reversible Boolean functions are called *cost-equivalent* if they belong to the same equivalence class (called *a cost-equivalence class*, or *a conjugacy class* in [2]) under simultaneous input/output relabeling and inversion.

It can be easily shown that two cost-equivalent functions have the same or reverse optimal circuits (i.e. optimal circuits for all cost-equivalent functions have the same cost for all cost metrics [2]). A cost-equivalence class can contain maximally $2{\cdot}n!$ functions (number of permutations of all variables doubled by the possibility of inversion).

We will be using the following popular vector notation for an *n*-variable reversible Boolean function *f*:

$$[f(0), f(1), \dots, f(2^n\text{-}1)],$$

where each binary vector *f(i)* will be expressed as a decimal.

TABLE I. TRUTH TABLE FOR FUNCTION $F$

| $x_3x_2x_1$ | $y_3y_2y_1$ |
|---|---|
| 0 0 0 | 0 0 1 |
| 0 0 1 | 0 0 0 |
| 0 1 0 | 0 1 1 |
| 0 1 1 | 0 1 0 |
| 1 0 0 | 1 0 1 |
| 1 0 1 | 1 1 1 |
| 1 1 0 | 1 0 0 |
| 1 1 1 | 1 1 0 |

**Definition 6.** The *canonical representative of a cost-equivalence class* is the function whose vector [$f$(0), $f$(1), ... , $f$($2^n$-1)] is lexicographically smallest.

**Example 1.** Let us consider a reversible function $f$ given in Table I. Its vector notation is [1,0,3,2,5,7,4,6]. It can be easily checked that this is the canonical representative of the cost-equivalence class of $f$ as all functions belonging to this class are as follows:

| | | |
|---|---|---|
| [1,0,3,2,5,7,4,6] | [1,0,3,2,6,4,7,5] | [1,0,3,7,5,4,2,6] |
| [1,0,6,2,5,4,7,3] | [2,3,0,1,5,7,4,6] | [2,3,0,1,6,4,7,5] |
| [2,3,0,7,6,1,4,5] | [2,5,0,1,6,7,4,3] | [4,3,6,7,0,1,2,5] |
| [4,5,3,7,0,1,2,6] | [4,5,6,1,0,7,2,3] | [4,5,6,2,0,1,7,3] |

## IV. A SURVEY OF RECENT RESULTS ON SYNTHESIS OF GATE COUNT OPTIMAL 3*3 AND 4*4 CIRCUITS

We have generated databases of *all* optimal circuits for 3*3 reversible Boolean functions for NCT library and two cost functions (GC and QC). After generating the databases we were using them in our research on synthesis of reversible circuits [6-9, 22-24]. During this research we have found all reversible functions requiring the maximal number of 8 gates in gate count optimal circuits. Next we discovered that for some of these functions longer circuits implementing them exist with much smaller QC. Fig. 1 shows such an example in which adding one gate leads to reducing QC from 24 to 13, i.e. by 45,8% (taking into account that two pairs of TOFFOLI and CNOT gates form two PERES gates). In general, this property is encountered in circuits of relatively large size. Thus, when looking for minimal quantum cost circuits one have to consider circuits of different sizes. This observation attracted our attention to the problem of minimizing QC in reversible circuits.

The databases generated by us for NCT library revealed one more interesting property of the set of all optimal 3*3 circuits. Namely, the greater the size of an optimal circuit the greater the number of all optimal circuits implementing the given function, e.g. the maximal number of all gate count as well as quantum cost optimal 3*3 circuits is higher than 1000 [8]. One can expect that in case of greater numbers of inputs/outputs the same property will hold and the maximal number of all gate count optimal 3*3 circuits will grow at least exponentially. Our recent experiments for 4-bit functions presented at the end of this paper show that this hypothesis is justified.

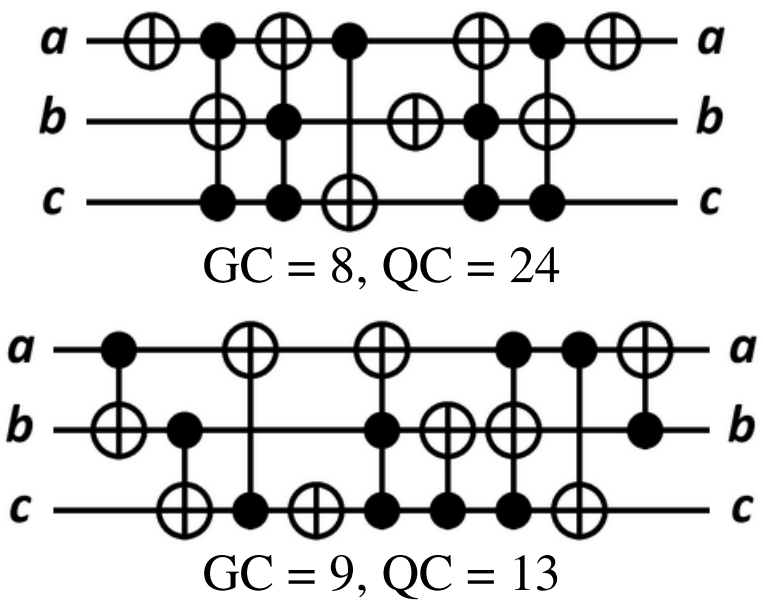

Figure 1. Two circuits implementing the same reversible function

TABLE II. DISTRIBUTION OF THE NUMBER OF CIRCUITS IMPLEMENTING 4*4 FUNCTION **NTH_PRIME4_INC** [17] OVER THEIR SIZE

| Gate count | # circuits |
|---|---|
| 11 | 12 |
| 12 | 2,288 |
| 13 | 187,945 |
| 14 | 11,056,332 |

Similar properties of reversible functions have been discovered by us with respect to the cardinality of the set of *all* (i.e. not only optimal) circuits implementing a given function. Table II shows how fast this number can grow for $n$ = 4 depending on the circuit size.

The above discussed properties demonstrate how difficult is the task of finding exact minimal quantum cost reversible circuits.

## V. GATE COUNT OPTIMAL CIRCUITS SYNTHESIS

In the years 2007-2009 the problem of developing algorithms for synthesis of all gate count optimal 4*4 reversible circuits using NCT library have attracted attention of some researchers [10, 11, 25, 29-31]. Only in one of them it was claimed that a hash table has been constructed consisting all 8-gate optimal circuits which potentially enables constructing optimal circuits up to size of 16 gates [10]. However, besides one relatively simple example no experimental data were supplied. In addition, the example showed that the tool was quite slow.

Recently, Golubitsky, Falconer and Maslov [2] presented the algorithm for finding a gate count optimal reversible circuit for any 4*4 reversible function. The algorithm is based on the fact that the set of all functions that have an optimal circuit up to 9 gates can be effectively stored in memory of today computers. It also relies on the fact that for each cost-equivalent class of reversible functions it is sufficient to store only its canonical representative, so the amount of required memory can be reduced almost 48 times . The functions in such database are stored in hash tables.

The database can be used directly to find GC of an optimal circuit for any reversible function $f$ up to 9 gates if the canonical representative of function $f$ is used for the database lookup.

Functions that require more than 9 gates in an optimal circuit need additional processing to determine GC of optimal circuits. Namely, the algorithm relies on the fact that any optimal circuit for the function $f$ can be partitioned into two circuits which realize functions $g$ and $r$, such as:

$$f = g \circ r,$$

where symbol ○ denotes composition (cascading of the circuits). The above equation can be transformed into the following:

$$f \circ r^{-1} = g.$$

The algorithm iterates over all functions in the database that have optimal circuits of length $i$ for $i = \{1,\ldots,9\}$, computes their inversions and then compose the function $f$ with each of them. For the resulting function $g$ it checks the length of an optimal circuit using the procedure from the beginning of the paragraph. If such function $g$ with optimal circuit of length $j$ has been found in the database, the length of the optimal circuit for the function $f$ equals to $i+j$ [2].

The above procedure provides an effective and very fast way of finding the length of an optimal circuit for any 4*4 reversible function. It can also be easily extended to a fast algorithm for finding an optimal circuit [2]. One just needs to store the last gate with each function in the database and the complete circuit can be easily constructed.

Using the computer system with 16 AMD Opteron 2300 MHz processors and 64GB RAM authors of [2] managed to create a database for 9-gate optimal circuits and synthesize an optimal circuit for a given 4*4 reversible function in about 0.01s on average.

A similar approach has been presented in [10]. However, the authors decided to store the complete circuits in the database. They only had managed to create a database for $n$=8. This decreased the synthesis speed so generating an optimal circuit for an example benchmark took approximately 35s.

## VI. Our synthesis algorithm

We have decided to extend the method presented in [2] for developing a tool capable of finding *all* reversible circuits with a specified gate count implementing the given reversible function. The algorithm is very similar to many heuristic algorithms of reversible circuit synthesis described in the literature. In each step a reversible gate is selected and added at the end of the current gate cascade (see Fig. 2). It might be necessary to backtrack and try another gate if the circuit for the given reversible function has not been found (Fig. 3). For selecting a gate algorithms rely on the criterion called a complexity measure. It limits the search tree and allows to avoid selecting gates that do not lead to an optimal circuit [28], [29].

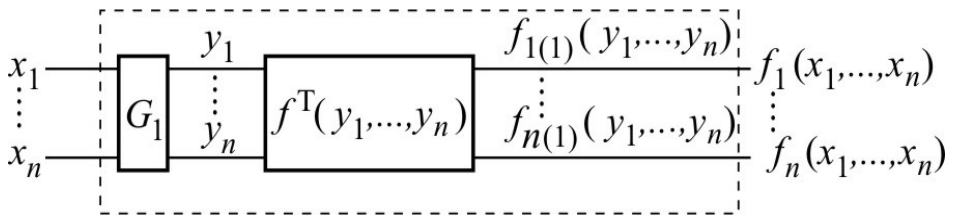

Figure 2. A general scheme of one step of a search-based algorithm

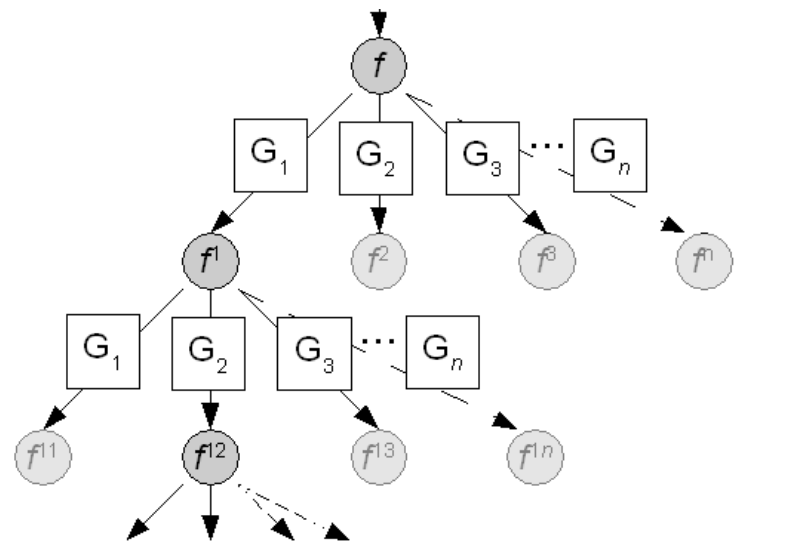

Figure 3. Example of the search tree of a search-based algorithm

**Definition 7.** *A partial realization* of a given reversible Boolean function $f$ is a gate cascade, such that after adding a gate (or gates) a circuit implementing $f$ is created.

**Definition 8.** *A remainder function* is a Boolean function to be implemented in a circuit that has to be added to the current partial realization to get a circuit implementing a specified reversible function.

In Fig. 2 $f^{\mathrm{T}}$ denotes the remainder function to be implemented after selecting $G$ as the first gate of the circuit under construction.

**Definition 9.** A *complexity measure* is a function that assigns a number to each reversible function in such a manner that functions with complex optimal circuits have higher measures than functions with simple optimal circuits. The *perfect complexity measure* assigns a monoto-nically decreasing numbers to remainder functions after adding consecutive gates in an optimal circuit for a given reversible function.

Note that the simple search-based algorithm is only guided by a complexity measure. It selects only the gates that generate decreasing complexity measure for the remainder functions. Even such simple algorithm can find more than one reversible circuit if it continues to run after finding the first reversible gate cascade that implements the specified function and use backtracking to check all paths in search tree.

## VII. Experimental results

It can be easily noticed that the gate count calculated with the algorithm from [2] is a perfect complexity measure for the heuristic algorithms. We have decided to perform an experiment for finding quantum cost optimal circuits with specified sizes for previously proposed 4*4 reversible benchmarks. As a result of our calculations of all optimal circuits for 3*3 reversible functions we expected in advance that in some cases the runtimes could be very high. The results of our calculations are presented in Tables III-VI. The reversible circuits that have been found with this approach improve substantially previously reported circuits with best known quantum cost. The minimal value of the quantum cost of circuits found by us is sometimes more than two times smaller than the maximal one. For some benchmarks improvement was up to 74%.

Our initial analysis of optimal circuits for 3*3 reversible functions showed that the quantum cost optimal circuits have higher average GC than the gate count optimal circuits. This is why we considered an extension to the algorithm for synthesis of all 4*4 reversible circuits. Namely, increasing by *a* the initial value considered as the complexity measure for function *f*, the algorithm will find all reversible circuits of length equal to *o*, *o*+1, …, *o*+*a*, where *o* is the length of the optimal circuit for the function *f*. The value of the parameter *a* needs to be adjusted experimentally to limit the calculation time and the number of generated circuits. Both the calculation time and the number of the circuits grow exponentially with the increasing value of *a* (see Tables III-VI).

The proposed approach allows to find the 4*4 reversible quantum cost optimal circuits for a specified length of the reversible gate cascade. This does not mean that these circuits are exact minimal in terms of quantum cost, because there might exist longer cascades with lower quantum cost value (see the results in Tables III-VI). Nevertheless, the circuits we have found using the above described method are usually better in terms of quantum cost than the best circuits known from the literature.

Our algorithm can be very easily parallelized. Calculations of the optimal circuit length for the remainder functions after applying all possible gates can be performed independently. In practical applications one should also consider adding a time limit, because the calculation take a lot of time for the parameter *a* larger than 2 and circuits longer than 13 gates. The calculation time also depends on the size of the database and the computer system used for calculations. Our results have been obtained on IBM pSeries p5 550 with two 4-way 1.65GHz POWER5+ CPUs with total of 16GB of memory. We used a database for 8-gate circuits. All the times are given in *cpu-seconds*.

To present our results in a compact way we have shortened the names of gates:

- NOT(a) is denoted by Na,
- CNOT(a,b) is denoted by Ca-b,
- TOFFOLI(a,b,c) is denoted by Tab-c,
- TOFFOLI(a,b,c,d) is denoted by T4abc-d.

We have adopted a method of calculating quantum cost applied in [12]. Possibility of grouping T and C gates into Peres gates or inverted Peres gates are detected and marked in the following way: <Tab-c Ca-b> or <Ca-b Tab-c>.

Our results are partitioned into four tables for four sets of benchmarks to which they are compared. For example, Table III compares our synthesis results with the ones reported for the five cost-equivalent classes of maximal size (**4g15b** functions) posted in [12] and obtained with the fast optimization tool described in [2]. Quantum cost of our circuits (in bold in Table III) is on average 33.1% off from the earlier reported, with 49.2% for **4b15g_2** (see Table VII). Our circuits were found by generating all 15-gate gate count minimal circuits thus present exact minimal values for circuits with 15 gates. The interval width of quantum cost values varies from 36 (for **4b15g_1**) up to 56 (for **4b15g_5**). The number of generated circuits varies from 79277 (for **4b15g_5**) up to 340066 (for **4b15g_1**). The circuits collected in the Table III are also showed graphically in Fig. 4. Some of the other circuits found by us are shown in Fig. 5 and Fig. 6 together with the source circuits.

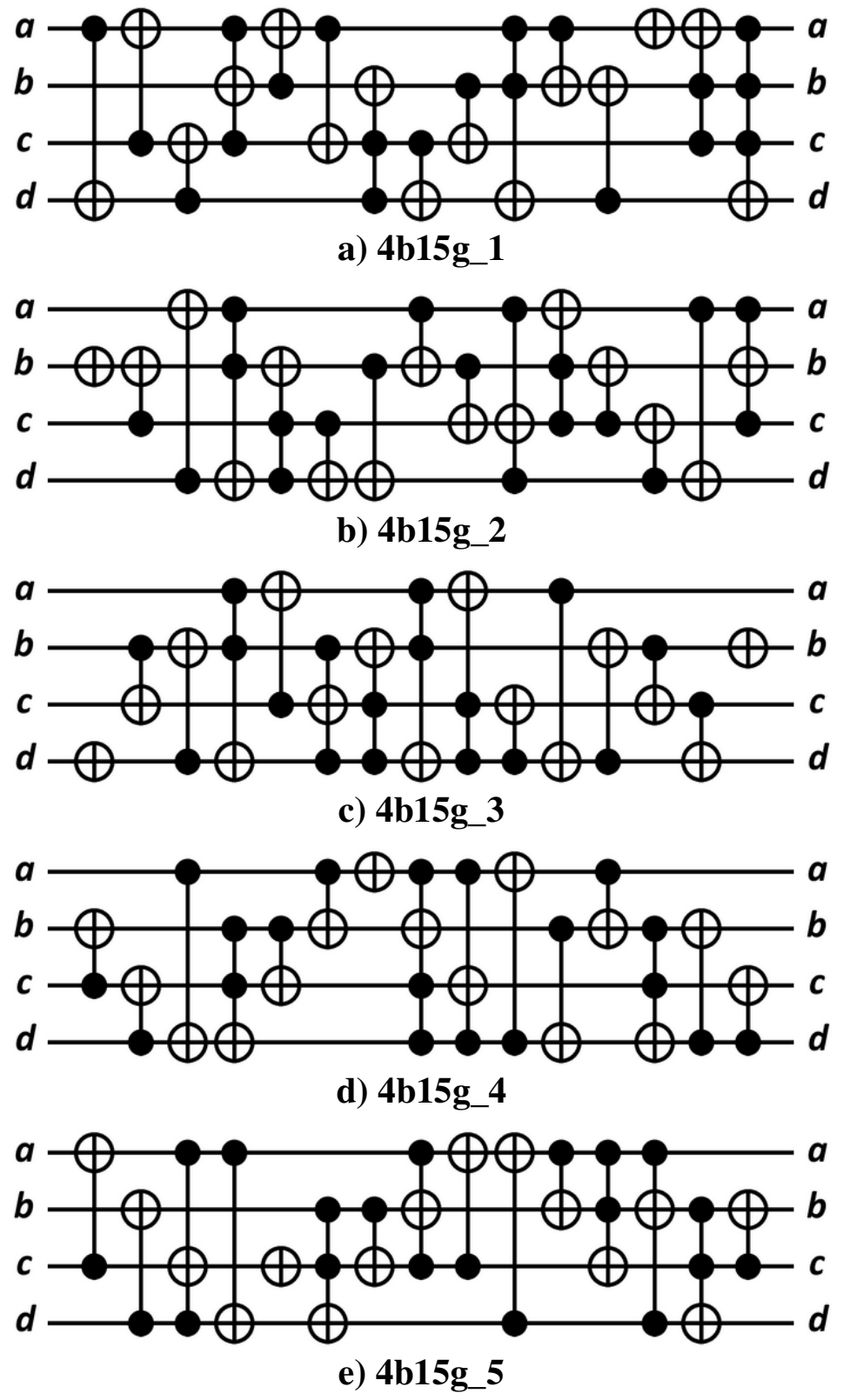

Figure 4. Our circuits with minimal quantum cost for **4b15g** functions

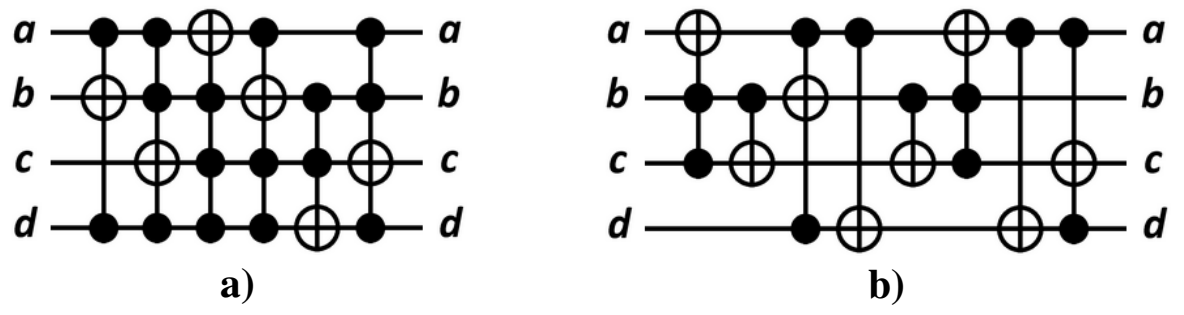

Figure 5. Circuits for the function **mini_alu**: a) best known from [26], b) best result of our optimization (Table V)

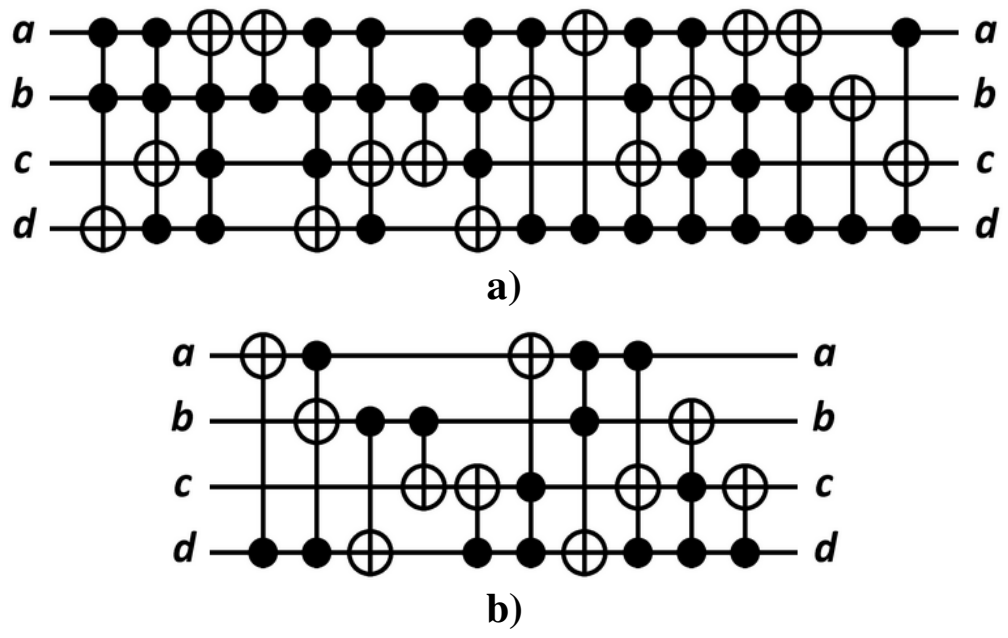

Figure 6. Circuits for the function **dmasl**: a) example from [15], b) best result of our optimization (Table V)

TABLE III. QUANTUM COST OPTIMAL IMPLEMENTATIONS WITH SPECIFIED GATE COUNT FOR **4G17G** BENCHMARK 4*4 FUNCTIONS [12]

| GC | QC | #circuits | time[s] | example circuit for $QC_{min}$ |
|---|---|---|---|---|
| **4b15g_1 [1,5,0,8,9,11,2,15,3,12,4,6,10,14,13,7] GC=15, QC=47** in [12] | | | | |
| 15 | **39**-75 | 340066 | 366283 | Ca-d Cc-a Cd-c Tac-b Cb-a Ca-c <Tcd-b Cc-d> Cb-c <Tab-d Ca-b> Cd-b Na Tbc-a T4abc-d |
| **4b15g_2 [1,9,0,4,10,8,2,11,3,15,5,12,7,14,13,6] GC=15, QC=61** in [12] | | | | |
| 15 | **31**-83 | 207520 | 395002 | Nb Cc-b Cd-a Tab-d <Tcd-b Cc-d> Cb-d Ca-b Cb-c Tad-c <Tbc-a Cc-b> Cd-c Ca-d Tac-b |
| **4b15g_3 [3,1,7,13,11,0,8,15,2,5,10,6,9,14,12,4] GC=15, QC=53** in [12] | | | | |
| 15 | **33**-83 | 119857 | 369687 | Nd Cb-c Cd-b Tab-d Cc-a Tbd-c Tcd-b Tab-d <Tcd-a Cd-c> Ca-d Cd-b Cb-c Cc-d Nb |
| **4b15g_4 [3,1,11,7,8,0,9,5,2,6,15,13,14,4,10,12] GC=15, QC=47** in [12] | | | | |
| 15 | **35**-79 | 247186 | 350715 | Cc-b Cd-c Ca-d <Tbc-d Cb-c> Ca-b Na T4acd-b <Tad-c Cd-a> Cb-d Ca-b Tbc-d Cd-b Cd-c |
| **4b15g_5 [3,5,11,1,8,0,9,7,2,6,14,13,10,4,12,15] GC=15, QC=43** in [12] | | | | |
| 15 | **31**-87 | 79277 | 300135 | Cc-a Cd-b <Tad-c Ca-d> Nc <Tbc-d Cb-c> <Tac-b Cc-a> Cd-a <Ca-b Tab-c> Tad-b <Tbc-d Cc-b> |

TABLE IV. QUANTUM COST OPTIMAL IMPLEMENTATIONS WITH SPECIFIED GATE COUNT FOR 4*4 BENCHMARK FUNCTIONS FROM [3, 2, 12]

| GC | QC | #circuits | time[s] | example circuit for $QC_{min}$ |
|---|---|---|---|---|
| **4_49 [15,1,12,3,5,6,8,7,0,10,13,9,2,4,14,11] GC=12, QC=32** in [3] | | | | |
| 12 | 30-72 | 374 | 73 | Cc-a Na Ca-d Tab-d Tcd-b <Cd-a Tad-c> Tbc-a Tab-d Cd-b Cd-c Na |
| 13 | 29-93 | 39121 | 2870 | Cc-a Ca-d Nd Tab-d Cd-b <Cb-c Tbc-d> <Cb-a Tab-c> Tcd-a Tad-b Cb-d Cd-c |
| 14 | **28**-104 | 2895738 | 118889 | Cb-d Cc-a Ca-d Nd Tab-d Tcd-b <Tad-c Cd-a> <Cb-c Tbc-a> <Tab-d Cb-a> Cd-b Cb-c |
| **decode42 [1,2,4,8,0,3,5,6,7,9,10,11,12,13,14,15] GC=10, QC=30** in [3] | | | | |
| 10 | **28**-30 | 16 | 2 | Cc-a Cc-b <Cd-a Tad-b> Cb-c T4abc-d Tbd-c Cc-a Ca-b Na |
| 11 | 29-51 | 1380 | 35 | Ca-b Cc-a Ca-b <Cd-a Tad-b> Cb-c T4abc-d Tbd-c Cc-a Ca-b Na |
| **hwb4 [0,2,4,12,8,5,9,11,1,6,10,13,3,14,7,15] GC=11, QC=21** in [3] | | | | |
| 11 | 21-39 | 264 | 18 | Ca-c <Cb-d Tbd-a> Cd-a Cc-d Tad-b Tbc-a Cd-c Cb-d Cc-b Ca-c |
| 12 | 20-60 | 38072 | 850 | Ca-b Cd-b Cb-c Cc-a <Ca-b Tab-d> Tcd-a Cb-c <Cc-a Tac-d> Cd-b Ca-c |
| 13 | **19**-67 | 1801004 | 33861 | Ca-b Cb-d Cd-c <Ca-d Tad-b> <Cc-b Tbc-a> Ca-b Cc-d Ca-c <Cd-a Tad-b> Cb-d |
| **imark [4,5,2,14,0,3,6,10,11,8,15,1,12,13,7,9] GC=7, QC=19** in [3] | | | | |
| 7 | 19-19 | 8 | 1 | Tcd-a Cb-c Tab-d Cd-a Cd-c Tac-b Nc |
| 8 | 18-44 | 303 | 6 | Tcd-a Ca-c Cb-c Nc Tab-d Cd-a <Tac-b Ca-c> |
| 9 | **17**-63 | 6437 | 108 | Cb-a Tcd-a Ca-c <Cb-a Tab-d> Cd-a <Ca-c Tac-b> Nc |
| **mperk [3,11,2,10,0,7,1,6,f,8,14,9,13,5,12,4] GC=9, QC=15** in [3] | | | | |
| 9 | **13**-17 | 76 | 2 | Nc <Cd-c Tcd-b> <Tac-d Cc-a> Cb-a Cd-a Ca-b Cb-c |
| 10 | 14-38 | 3411 | 49 | Ca-b Cd-c Nc <Tac-d Cc-a> Cb-a Ca-b Cd-b <Cb-c Tbc-a> |
| **oc5 [6,0,12,15,7,1,5,2,4,10,13,3,11,8,14,9] GC=11, QC=39** in [3] | | | | |
| 11 | 39-39 | 48 | 11 | Tbd-c Tcd-b Ca-c Tab-c Cc-a Cd-b Nc Tbc-d Ca-c Cc-b T4abd-c |
| 12 | **34**-64 | 8844 | 235 | Cb-c Na <Tcd-b Cd-c> Ca-d <Cd-b Tbd-c> Cc-a Tbc-d <Ca-c Tac-d> T4abd-c |
| 13 | 35-85 | 471283 | 9576 | Cb-c Cc-a Na <Tcd-b Cd-c> Ca-c Tbc-a Cd-b Cd-c Tab-d Cc-a Cc-b T4abd-c |
| **oc6 [9,0,2,15,11,6,7,8,14,3,4,13,5,1,12,10] GC=12, QC=42** in [3] | | | | |
| 12 | 38-68 | 4475 | 173 | Cc-a Cd-c Tbc-d Tad-b T4abd-c <Tbc-a Cc-b> Cb-d Tad-c Na Ca-d Cb-a |
| 13 | **37**-89 | 328105 | 8341 | Cb-a T4abc-d <Cd-b Tbd-a> Cc-b Na Ca-d Tab-d <Tcd-a Cd-c> Tbc-d Ca-c Cc-d |
| **oc7 [6,15,9,5,13,12,3,7,2,10,1,11,0,14,4,8] GC=13, QC=41** in [3] | | | | |
| 13 | 35-69 | 1423 | 970 | <Cb-d Tbd-a> Tab-c Cc-a Cd-c <Tad-b Ca-d> Cb-a Nb Tabc-d <Cc-d Tcd-b> Cb-c |
| 14 | **34**-90 | 201317 | 35102 | <Tbd-a Cb-d> <Ca-b Tab-c> <Cc-d Tcd-a> Cb-a Cd-b Tabc-d <Tbd-c Cb-d> Ca-c Nc Cc-b |
| **oc8 [11,3,9,2,7,13,15,14,8,1,4,10,0,12,6,5] GC=12, QC=48** in [3] | | | | |
| 12 | 40-48 | 473 | 64 | Ca-b Na Tcd-b T4abd-c Cb-d Tad-b Cb-a Ca-c Nb Tcd-b Tbc-d Ca-c |
| 13 | **35**-69 | 51771 | 2616 | <Cd-c Tcd-b> <Ca-b Tab-d> Nd Cd-a Cd-b T4abd-c Cb-d <Tcd-b Cd-c> Tbc-d Ca-c |
| **nth_prime4_inc [0,2,3,5,7,11,13,1,4,6,8,9,10,12,14,15] GC=15, QC=51** in [12] | | | | |
| 11 | 53-55 | 12 | 10 | Tab-c Tac-b <Cd-b Tbd-c> Tbc-d Ca-b T4bcd-a Cc-b Cb-a Tbd-a T4abd-c |
| 12 | 32-46 | 2288 | 591 | Cd-b <Cb-c Tbc-d> Ca-b Tad-b Cc-a <Cb-a Tab-c> Tcd-a Tbd-c Tab-d Cc-b |
| 13 | 31-93 | 187945 | 7282 | Cd-b Ca-d <Cc-b Tbc-d> Tad-c <Cb-a Tab-c> Tcd-b Ca-d Tbd-a <Cc-a Tac-d> Cb-c |
| 14 | **26**-114 | 11056332 | 292578 | Cd-b <Cc-b Tbc-d> Cb-d <Cd-a Tad-c> Cb-a <Ca-c Tac-b> Cd-a Tbd-a <Cc-a Tac-d> Cb-c |
| **primes4 [2,3,5,7,11,13,0,1,4,6,8,9,10,12,14,15] GC=10, QC=42** in [2] | | | | |
| 10 | 42-48 | 7 | 2 | Cd-c Cc-a Cb-c Nb Tbc-d T4abd-c Tac-b Na T4acd-b Cb-a |
| 11 | 23-71 | 8552 | 147 | Cd-a <Tcd-b Cc-d> Cb-d Tad-b <Cb-c Tbc-d> Cc-a Cd-c Nb Tbc-d |
| 12 | **22**-92 | 430950 | 6789 | Cb-c Cd-c <Cc-b Tbc-d> <Ca-b Tab-c> Cd-b Nb Tbc-a Cd-b <Ca-b Tab-c> |

TABLE V. QUANTUM COST OPTIMAL IMPLEMENTATIONS WITH SPECIFIED GATE COUNT FOR 4*4 FUNCTIONS FROM [26, 10, 19, 31, 15, 32]

| GC | QC | #circuits | time[s] | example circuit for $QC_{min}$ |
|---|---|---|---|---|
| **mini_alu [0,1,2,3,4,5,14,11,8,6,10,9,12,15,13,7] GC=6, QC=62** in [26] | | | | |
| 6 | 30-62 | 12 | 1 | Tbc-a Tad-b Tad-c Tbc-a Tbc-d Tad-c |
| 7 | 23-67 | 108 | 9 | <Tbc-a Cb-c> Tad-b <Cb-c Tbc-a> Tbc-d Tad-c |
| 8 | **16**-88 | 2795 | 109 | <Tbc-a Cb-c> <Tad-b Ca-d> <Cb-c Tbc-a> <Ca-d Tad-c> |
| 9 | 17-93 | 41586 | 755 | <Cb-c Tbc-a> Cb-a <Tad-b Ca-d> <Cb-c Tbc-a> <Ca-d Tad-c> |
| 10 | 18-114 | 880918 | 15216 | Cb-a <Cb-c Tbc-a> Ca-b <Ca-d Tad-b> <Cb-c Tbc-a> <Ca-d Tad-c> |
| **aj-e11 [1,2,4,8,0,3,5,6,7,9,10,11,12,13,14,15] GC=10, QC=30** in [26] | | | | |
| 10 | **28**-30 | 16 | 2 | Cc-a Cc-b <Cd-a Tad-b> Cb-c T4abc-d Tbd-c Cc-a Ca-b Na |
| 11 | 29-51 | 1380 | 35 | Ca-b Cc-a Ca-b <Cd-a Tad-b> Cb-c T4abc-d Tbd-c Cc-a Ca-b Na |
| 12 | 28-64 | 93818 | 1618 | Cc-a Cc-b Cb-c <Cd-b Tbd-c> <Tad-b Cd-a> T4abc-d Cd-c Cc-a Ca-b Na |
| 13 | 29-85 | 4992518 | 78170 | Ca-b Ca-c <Cd-b Tbd-a> <Cc-d Tcd-b> T4abd-c Cb-a Cc-d Ca-c Cd-a Ca-b Na |
| **mod10_171 [1,2,3,4,5,6,7,8,9,0,10,11,12,13,14,15] GC=10, QC=56** in [26] | | | | |
| 9 | 53-61 | 132 | 5 | <Tbd-a Cd-b> T4bcd-a T4abd-c T4abc-d <Tab-c Ca-b> Na Tad-b |
| 10 | 38-66 | 6856 | 129 | Cb-d T4abd-c Tac-b <Cb-d Tbd-a> <Ca-d Tad-b> Tcd-a Tab-d Na |
| 11 | 35-87 | 243918 | 4172 | Ca-c <Cd-c Tcd-a> Nc Tad-b T4abc-d <Tab-c Ca-b> Na Tcd-a Ca-c |
| 12 | **32**-100 | 7487866 | 120082 | <Cd-c Tcd-b> T4abc-d Ca-c <Tcd-a Cd-c> <Tab-c Ca-b> <Tcd-a Cd-c> Ca-c Na |
| **mod10_176 [1,2,3,4,5,6,7,8,9,0,11,12,13,14,15,10] GC=7, QC=41** in [26] | | | | |
| 7 | 35-43 | 29 | 1 | Ca-c Tab-c T4acd-b T4abc-d Ca-b Ca-c Na |
| 8 | 30-64 | 675 | 12 | Tbc-d Tab-c Tad-b Tbc-d <Tab-d Ca-b> Tad-b Na |
| 9 | **21**-69 | 20486 | 311 | Cd-c Tab-d <Tcd-b Cd-c> <Tab-d Ca-b> Cd-b Na Tcd-b |
| 10 | 22-90 | 86848 | 1298 | Ca-c Cc-d Tab-c <Cc-d Tcd-b> <Tab-d Ca-b> Tcd-b Ca-c Na |
| **4_49+hwb4 [15,2,3,12,5,9,1,11,0,10,14,6,4,8,7,13] GC=12, QC=30** in [10] | | | | |
| 12 | 28-34 | 22 | 47 | Cc-b Cb-d Nd Tad-b <Tcd-a Cd-c> Ca-d Tab-c <Tbc-a Cc-b> Tad-b Cb-a |
| 13 | 27-65 | 6847 | 1479 | Cc-b Cb-d Ca-b Nd <Tcd-a Cd-c> <Ca-d Tad-b> Tab-c <Tbc-a Cc-b> Tad-b Cb-a |
| 14 | **26**-94 | 50442 | 670503 | Cc-b Cb-d Ca-b Nd <Tcd-a Cd-c> <Tad-b Ca-d> <Ca-b Tab-c> <Tbc-a Cc-b> Tad-b Cb-a |
| **msaee [11,3,9,2,7,13,15,14,8,1,4,10,0,12,6,5] GC=16, QC=72** in [19] | | | | |
| 12 | 40-48 | 473 | 65 | Ca-b Na Tcd-b T4abd-c Cb-d Tad-b Cb-a Ca-c Nb Tcd-b Tbc-d Ca-c |
| 13 | 35-69 | 51771 | 2630 | <Cd-c Tcd-b> <Ca-b Tab-d> Nd Cd-a Cd-b T4abd-c Cb-d <Tcd-b Cd-c> Tbc-d Ca-c |
| 14 | **34**-90 | 2319018 | 81388 | <Cd-c Tcd-b> <Ca-b Tab-d> Nd Cd-a <Tbc-d Cb-c> Cd-b Cb-c T4abd-c <Tbc-d Cb-c> Ca-c |
| **gyang [2,5,3,15,4,13,6,7,8,9,10,11,12,1,14,0] GC=19, QC=103** in [31] | | | | |
| 10 | 52-60 | 156 | 4 | Na <Ca-c Tac-d> T4bcd-a T4acd-b Cc-d Na Nc Tad-c T4abc-d |
| 11 | 37-67 | 8867 | 165 | Nc Tac-d <Cc-d Tcd-a> Tab-c T4acd-b <Tcd-a Cc-d> Nc <Tad-c Ca-d> |
| 12 | **36**-88 | 372942 | 6161 | Nc <Tac-d Cc-a> <Tcd-a Cc-d> Tab-c T4acd-b <Tcd-a Cc-d> Nc <Tad-c Ca-d> |
| **dmasl [0,1,14,3,4,5,7,8,15,13,10,6,9,12,11,2] GC=16, QC=128** in [15] | | | | |
| 9 | 25-49 | 65 | 3 | <Cd-a Tad-b> Cd-c Cb-d Tcd-a Tab-d Tad-c <Tcd-b Cd-c> |
| 10 | **24**-64 | 2269 | 40 | <Cd-a Tad-b> Cb-d Cb-c <Cd-c Tcd-a> Tab-d Tad-c <Tcd-b Cd-c> |
| 11 | 25-75 | 84522 | 1599 | <Cd-a Tad-b> Cd-c Cb-d Tcd-a Tab-d <Tad-c Cd-a> <Tcd-b Cd-c> Cd-a |
| 12 | **24**-96 | 2637559 | 42517 | <Cd-a Tad-b> Cb-d Cb-c <Cd-c Tcd-a> Tab-d <Tad-c Cd-a> <Tcd-b Cd-c> Cd-a |
| **App2.2 [7,14,9,6,11,0,13,2,5,15,10,12,1,4,3,8] GC=18, QC=102** in [32] | | | | |
| 11 | 39-71 | 101 | 8 | Tad-b Nc Tbcd-a Tab-c Nd <Tbd-c Cd-b> Tbc-d <Tad-b Ca-d> Na |
| 12 | 36-88 | 16532 | 443 | Ca-b Nc Tbcd-a Nd Cd-a Tab-c <Tad-b Cd-a> Tbc-d <Tad-b Ca-d> Na |
| 13 | **35**-109 | 1067635 | 21905 | Cd-c <Tad-b Cd-a> Tbcd-a Tab-c Cd-a Nd Cd-b <Cb-c Tbc-d> <Tad-b Ca-d> Na |
| **App2.11 [7,14,9,6,11,0,13,2,5,15,10,12,1,4,3,8] GC=14, QC=82** in [32] | | | | |
| 9 | 45-53 | 55 | 1 | Tabd-c Tac-b Cb-c Tabc-d Tbd-a Tad-b Cb-c Cc-d Nd |
| 10 | 30-74 | 2999 | 63 | Ca-c Tbc-d Tad-c Tbc-d Tac-b Tbd-a Ca-c Cc-d Ca-b Nd |
| 11 | 27-95 | 99722 | 1804 | <Ca-c Tac-b> Tbd-a <Tac-d Ca-c> Ca-b Tbd-a Na Tac-d Ca-d Na |
| 12 | **26**-112 | 2974831 | 53529 | <Ca-c Tac-b> Tbd-a Ca-b <Tac-d Cc-a> Tbd-c Nc Cc-a <Tac-d Ca-c> Na |

TABLE VI. QUANTUM COST OPTIMAL IMPLEMENTATIONS WITH SPECIFIED GATE COUNT FOR 4*4 FUNCTIONS FROM [30]

| GC | QC | #circuits | time[s] | example circuit for $QC_{min}$ |
|---|---|---|---|---|
| **f1 [0,1,2,3,15,10,11,13,9,12,5,4,14,8,6,7] GC=13, QC=24** in [30] | | | | |
| 9 | 27-33 | 12 | 1 | <Tcd-b Cc-d> Tbd-c Tbc-d Tad-c <Tbc-a Cc-b> <Tad-b Cd-a> |
| 10 | 26-70 | 680 | 16 | Cc-d <Tbd-c Cd-b> Tbc-d Cc-d Tad-c Tbc-a <Cd-a Tad-b> Cc-a |
| 11 | 23-91 | 34818 | 939 | Cc-b <Tbd-c Cd-b> Cc-d Cb-a <Cd-a Tad-c> Tbc-a Cb-a Tad-b Cc-a |
| 12 | **22**-100 | 1319130 | 24288 | Cb-c <Tcd-b Cc-d> Cd-a Cb-d Tad-c Cd-b <Tbc-a Cb-c> <Cd-a Tad-b> Cc-a |
| **f2 [0,1,2,3,15,10,11,14,8,7,4,13,6,9,5,12] GC=14, QC=30** in [30] | | | | |
| 8 | 46-48 | 11 | 3 | Tbcd-a Tad-c Cc-d Tacd-b Tcd-a Tbd-c <Tbc-d Cc-b> |
| 9 | 43-69 | 335 | 7 | Tbcd-a Ca-b Tbd-c <Cc-d Tcd-a> Tacd-b Ca-b Tbd-c Cc-b |
| 10 | 26-76 | 11511 | 217 | Cc-b Ca-b Tbd-a Tac-b Tbd-c Cc-d Cc-a Ca-b Tbd-c Cc-b |
| 11 | **25**-95 | 297276 | 5205 | Cc-a Ca-b Tbd-a <Cc-a Tac-b> Tbd-c Cc-d Cc-a Ca-b Tbd-c Cc-b |
| **g1 [0,1,2,3,4,5,6,7,8,9,13,12,14,15,11,10] GC=7, QC=11** in [30] | | | | |
| 3 | 15-15 | 2 | 0 | Tbd-c Tbd-a Tcd-b |
| 4 | 12-36 | 13 | 0 | Cc-a Tbd-c Tcd-b Cc-a |
| 5 | 13-41 | 150 | 4 | <Tbd-a <Cd-b> Tbd-c> <Tcd-b Cd-c> |
| 6 | **10**-62 | 1672 | 26 | Cc-a <Cd-b Tbd-c> <Tcd-b Cd-c> Cc-a |
| **g2 [0,1,2,3,4,5,6,7,8,10,12,14,15,13,11,9] GC=8, QC=12** in [30] | | | | |
| 4 | 20-20 | 2 | 0 | Tbd-c Tad-b Tcd-b Tbd-a |
| 5 | 17-25 | 18 | 0 | Tbd-c Cc-a Tad-b Tbd-a Cc-a |
| 6 | 14-46 | 449 | 10 | Cc-a Tbd-c Cb-a Tad-b Cc-a Cb-a |
| 7 | 15-67 | 5242 | 79 | Cb-c Cc-a Tad-b Tbd-c Cb-c Cc-a Cb-a |
| 8 | **12**-72 | 93966 | 1396 | Cc-a <Tbd-c Cd-b> Cb-a <Tad-b Cd-a> Cc-a Cb-a |
| **g3 [0,1,2,3,4,5,6,7,11,9,15,13,10,8,14,12] GC=8, QC=12** in [30] | | | | |
| 7 | 15-31 | 358 | 10 | Cc-b Tbd-c Cc-b Cb-a Tad-b Cd-b Cb-a |
| 8 | **12**-52 | 13944 | 183 | Cc-b <Tbd-c Cd-b> Cc-b Cb-a <Cd-a Tad-b> Cb-a |
| 9 | 13-73 | 313370 | 4680 | Cc-b <Tbd-c Cd-b> Cc-b Cb-a Cd-b <Tad-b Cd-a> Cb-a |
| **g4 [0,1,2,3,13,5,15,7,6,4,11,9,14,12,10,8] GC=12, QC=18** in [30] | | | | |
| 9 | 17-37 | 149 | 5 | Cb-a Tad-b Cd-b Cb-a <Tac-d Cc-a> <Tad-c Cd-a> Cc-d |
| 10 | **16**-58 | 13086 | 180 | Cb-a <Cd-a Tad-b> Cd-c Cb-a <Cc-a Tac-d> <Cd-a Tad-c> Cc-a |
| 11 | 17-79 | 485674 | 7034 | Cb-a <Cd-a Tad-b> Cd-c Cb-a <Cc-a Tac-d> Cd-c <Tad-c Cd-a> Cc-a |
| **g5 [0,1,2,3,12,5,14,7,6,4,10,8,11,9,15,13] GC=10, QC=19** in [30] | | | | |
| 8 | 20-32 | 54 | 6 | Cb-a <Cd-b Tbd-c> Tad-b <Tcd-a Cc-d> Cb-a Tac-d |
| 9 | 19-53 | 4744 | 68 | Ca-b <Tbd-a Cd-b> Tac-d Ca-b <Cd-a Tad-c> <Tcd-a Cc-d> |
| 10 | **18**-74 | 138151 | 2566 | <Cd-b Tbd-c> Cb-a <Cd-a Tad-b> <Cc-d Tcd-a> Cb-a <Cc-a Tac-d> |
| 11 | 19-91 | 3772042 | 61128 | Cb-a <Tbd-c Cd-b> <Tad-b Cd-a> <Tcd-a Cd-c> Cc-a Cb-a <Tac-d Cc-a> |
| **g6 [0,1,2,3,13,5,15,7,6,4,11,9,10,8,14,12] GC=11, QC=19** in [30] | | | | |
| 8 | 32-32 | 8 | 1 | Tad-b Tbd-a Na Tad-c Tad-b Tac-d Tcd-a Na |
| 9 | 19-53 | 1863 | 24 | Ca-b <Cd-b Tbd-a> <Tad-c Cd-a> Tac-d Ca-b <Cc-d Tcd-a> |
| 10 | **18**-70 | 48079 | 931 | Cb-a <Cd-b Tbd-c> <Tad-b Cd-a> Cb-a <Tac-d Cc-a> <Tcd-a Cc-d> |
| 11 | 19-91 | 1391436 | 25668 | Cb-a <Cd-b Tbd-c> <Tad-b Cd-a> Cb-a <Tac-d Cc-a> <Cc-d Tcd-a> Cc-a |
| **g7 [0,1,2,3,4,13,6,15,14,12,10,8,11,9,7,5] GC=11, QC=19** in [30] | | | | |
| 7 | 21-31 | 26 | 2 | Cb-a <Cd-b Tbd-c> Tad-b Tcd-a Cb-a Tac-d |
| 8 | 20-52 | 1474 | 21 | Tbd-c <Tcd-b Cd-c> Ca-b <Tbd-a Cd-b> Tac-d Ca-b |
| 9 | 19-73 | 34322 | 904 | Cb-a <Cd-b Tbd-c> Tad-b <Cc-d Tcd-a> Cb-a <Tac-d Cc-a> |
| 10 | **18**-90 | 812750 | 15728 | <Cd-b Tbd-c> Cb-a <Cd-a Tad-b> <Cc-d Tcd-a> Cb-a <Tac-d Cc-a> |
| **g8 [0,1,2,3,14,5,13,7,6,4,9,10,8,11,12,15] GC=13, QC=23** in [30] | | | | |
| 9 | 31-51 | 87 | 5 | Tad-b Tbd-a Tac-d Cd-c Cc-b Tad-c <Cc-d Tcd-a> Tbd-a |
| 10 | 24-70 | 5634 | 132 | Cb-a <Cd-b Tbd-c> Tad-b <Tcd-a Cc-d> Cb-a Tac-d <Tbd-a Cd-b> |
| 11 | **21**-91 | 222473 | 3660 | Tad-c Cb-a Cc-b <Cd-a Tad-c> Cb-a <Tac-d Cc-a> <Cd-b Tbd-a> Cc-b |
| **g9 [4,5,6,7,8,1,10,3,2,0,14,12,15,13,11,9] GC=12, QC=20** in [30] | | | | |
| 9 | 21-33 | 72 | 4 | Cb-a <Cd-b Tbd-c> Tad-b <Tcd-a Cc-d> Cb-a Tac-d Nc |
| 10 | 20-54 | 9847 | 221 | Ca-b <Tbd-a Cd-b> Tac-d Ca-b <Cd-a Tad-c> <Tcd-a Cc-d> Nc |
| 11 | **19**-75 | 345461 | 4969 | <Cd-b Tbd-c> Cb-a <Cd-a Tad-b> <Cc-d Tcd-a> Cb-a <Cc-a Tac-d> Nc |

TABLE VI. (CONTINUATION) QUANTUM COST OPTIMAL IMPLEMENTATIONS WITH SPECIFIED GATE COUNT FOR 4*4 FUNCTIONS FROM [30]

| GC | QC | #circuits | time[s] | example circuit for $QC_{min}$ |
|---|---|---|---|---|
| **g10 [0,1,10,11,12,5,7,15,14,4,3,8,2,9,6,13] GC=13, QC=23** in [30] | | | | |
| 9 | 27-27 | 1 | 1 | Cb-a <Cd-b Tbd-c> Tad-b Cb-d Tcd-a Tac-d Tbd-a Cc-d |
| 10 | 24-52 | 385 | 13 | Cb-a <Cd-b Tbd-c> Tad-b <Tcd-a Cc-d> Cb-a Tac-d <Tbd-a Cb-d> |
| 11 | 23-75 | 35538 | 839 | Ca-b <Tbd-a Cd-b> Tac-d Ca-b <Cd-a Tad-c> <Tcd-a Cc-d> <Tbd-a Cb-d> |
| 12 | **22**-92 | 1473728 | 29296 | <Cd-b Tbd-c> Cb-a <Cd-a Tad-b> <Cc-d Tcd-a> Cb-a <Cc-a Tac-d> <Tbd-a Cb-d> |
| **g11 [0,1,2,3,12,9,8,15,11,14,5,4,13,10,6,7] GC=16, QC=30** in [30] | | | | |
| 9 | 41-53 | 112 | 5 | Tbcd-a Cd-a <Tbc-d Cc-b> Tacd-b Cd-c <Tbc-d Cc-b> Tad-c |
| 10 | 30-70 | 3883 | 112 | Tcd-b <Cb-a Tab-d> Tcd-a Tab-d <Tbd-c Cd-b> Cc-d Tad-c Cb-a |
| 11 | **25**-87 | 130084 | 2463 | Cc-b <Tbd-c Cd-b> Cc-d Cb-a Tad-c Tbc-a Tad-b Cd-c Cc-a Cb-a |
| **g12 [0,1,3,10,6,11,2,15,8,14,4,13,7,9,5,12] GC=18, QC=34** in [30] | | | | |
| 9 | 51-53 | 5 | 1 | Tbcd-a Tad-c <Cb-c Tbc-d> Tcd-a Tbd-c Cc-b Tab-d Tabd-c |
| 10 | 46-74 | 239 | 11 | Cb-c Tbcd-a <Cd-a Tad-c> Tbc-d <Tcd-a Cd-c> Cc-b Tab-d Tabd-c |
| 11 | 45-79 | 10503 | 350 | Tcd-a Tad-c <Cb-c Tbc-d> Tcd-a Tad-c Tac-b Tab-d Tbd-c Tac-b Cc-b |
| 12 | **28**-100 | 4072880 | 7560 | Tbc-a <Cd-a Tad-c> <Tbc-d Cc-b> Cd-b Tab-d <Tcd-a Cd-c> <Tab-c Cb-a> Cc-a |

TABLE VII. OUR IMPROVEMENTS OF QUANTUM COST

| Benchmark | best known | source | our best circuit | ΔQC | % impr. |
|---|---|---|---|---|---|
| **4b15g_1** | 47 | [3] | 39 | -8 | 17.0 |
| **4b15g_2** | 61 | [3] | 31 | -30 | 49.2 |
| **4b15g_3** | 53 | [3] | 33 | -20 | 37.7 |
| **4b15g_4** | 47 | [3] | 35 | -12 | 25.5 |
| **4b15g_5** | 43 | [3] | 31 | -12 | 27.9 |
| **nth_prime4_inc** | 51 | [3] | 26 | -25 | 49.0 |
| **4_49** | 32 | [3,12] | 28 | -4 | 12.5 |
| **decode42** | 30 | [3] | 28 | -2 | 6.7 |
| **hwb4** | 21 | [3,12] | 19 | -2 | 9.5 |
| **imark** | 19 | [3] | 17 | -2 | 10.5 |
| **mperk** | 15 | [3] | 13 | -2 | 13.3 |
| **oc5** | 39 | [3] | 34 | -5 | 12.8 |
| **oc6** | 42 | [3] | 37 | -5 | 11.9 |
| **oc7** | 41 | [3] | 34 | -7 | 17.1 |
| **oc8** | 48 | [3] | 35 | -13 | 27.1 |
| **primes4** | 42 | [3] | 22 | -20 | 47.6 |
| **mini_alu** | 62 | [26] | 16 | -46 | 74.2 |
| **aj_e11** | 30 | [26] | 28 | -2 | 6.7 |
| **mod10_171** | 56 | [26] | 32 | -24 | 42.9 |
| **mod10_176** | 41 | [26] | 21 | -20 | 48.8 |
| **4_49+hwb4** | 30 | [10] | 26 | -4 | 13.3 |
| **msaee** | 72 | [19] | 34 | -38 | 52.8 |
| **gyang** | 103 | [31] | 36 | -67 | 65.0 |
| **dmasl** | 128 | [15] | 24 | -104 | 81.3 |
| **App2.2** | 102 | [32] | 35 | -67 | 65.7 |
| **App2.11** | 82 | [32] | 26 | -56 | 68.3 |
| **average:** | 51.4 | | 28.5 | -23.0 | 44.7 |

Table IV compares our synthesis results to the ones collected in [2, 3] for a set of known benchmarks (we have omitted simple benchmarks **4-bit-7-8**, **rd32** and **shift4** as exact minimal quantum cost circuits for them have already been found). The comparison is favorable for our circuits, (e. g., our circuits for the functions **nth_prime4_inc** and **primes4** are approximately 50% off from those reported in [3]). Again, the lowest values of quantum cost for circuits found by us are shown in bold in the column QC. Note that usually initial increasing of circuit size leads to lowering quantum cost but further increasing of GC causes a rise of QC (see results for benchmarks **imark** and **oc5**).

Table V compares our synthesis results to 4*4 designs we could find in RevLib [26] and in recent publications. For some of these functions (e.g. **mini-alu** and **gyang**) improvements (shown in bold) of over 60% are reported. Table VI compares the results of our synthesis algorithm with the results in [30] where 4*4 quantum cost optimal circuits built from Peres gates were searched. For a fair comparison we calculated GC in circuits from [30] by considering a Peres gate as two gates like in our designs. For 12 out of 14 functions considered in [30] our designs have smaller QC and for the other two functions we have obtained the same minimal values of QC.

Table VII shows the collective data comparing our circuits to previously found ones realized by using gates exclusively from NCT library. These data were taken from Tables III-V. For some benchmarks and designs taken from the literature we have obtained savings in QC of over 74% comparing with previously known circuits. Average improvement is equal to 44.7 %. In Fig. 5 and Fig. 6 circuits for our best QC improvement (**mini_alu** [26] and **dmasl** [15]) are shown together with the source circuits found in the literature. For some of the benchmarks considered in the paper better results could probably be generated, however we had to stop calculations due to time limitations in our experiments.

Studying Tables III-VI one can see that increasing the gate count up to a certain value causes the reduction of quantum cost of best circuits. Usually this quantum cost reduction is achieved by replacing pairs of multiple-control Toffoli gates by cascades built from smaller gates. Further reduction of QC is achieved by grouping Toffoli and CNOT gates into Peres gates. The presented method does not guarantee finding quantum cost minimal circuits because the length (GC) of such circuits is not known. However, for an arbitrary selected circuit length it allows for generating 4*4 circuits with the least quantum cost if the computation time is not too large. Due to this property we were able to find circuits with QC smaller than results obtained by the other methods presented in the literature.

## VIII. Conclusions and future work

A common practice in synthesis of reduced quantum cost reversible circuits (e.g. see [5], page 704) is first find a gate count optimal circuit and then map the resulting circuit into quantum gates. However, as we showed in the paper, this approach does not lead to minimal quantum cost circuits.

It requires considering circuits having greater number of gates than the minimal one to be able to find exact minimal quantum cost circuits. Research presented in the paper can be extended in different directions. One possibility is to consider other extensions to quantum cost model, like for example inter-gate optimizations [16, 17, 20, 21]. The other possibility is to apply the proposed method for another criteria of optimality. Namely, the large number of the reversible circuits that realize the same reversible function suggest that the circuits might also differ significantly under other criteria, e.g. depth of a circuit, latency, testability properties or new cost functions (like LNN cost or transistor cost [28]).

Due to limitations in speed and memory space of current computer systems the proposed method cannot be used directly for circuits with more than 4 inputs/outputs. However, it might be applied to optimize some parts of $n*n$ circuits, $n> 4$, having relatively long 4*4 subcircuits.

## Acknowledgment

This work was supported by the Polish Ministry of Science and Higher Education under Grant 4180/B/T02/2010/38.